\documentclass[sigconf]{acmart}

\usepackage[frozencache=true,cachedir=minted-cache]{minted} 
\usepackage[compatibility=false]{caption}

\copyrightyear{2022}
\acmYear{2022}
\setcopyright{acmlicensed}\acmConference[CHI '22]{CHI Conference on Human Factors in Computing Systems}{April 29-May 5, 2022}{New Orleans, LA, USA}
\acmBooktitle{CHI Conference on Human Factors in Computing Systems (CHI '22), April 29-May 5, 2022, New Orleans, LA, USA}
\acmPrice{15.00}
\acmDOI{10.1145/3491102.3502004}
\acmISBN{978-1-4503-9157-3/22/04}

\begin{document}

\title{Jury Learning: Integrating Dissenting Voices into Machine Learning Models}




\author{Mitchell L. Gordon}
\affiliation{%
  \institution{Stanford University}
  \city{Stanford}
  \country{USA}}
\email{mgord@cs.stanford.edu}

\author{Michelle S. Lam}
\affiliation{%
  \institution{Stanford University}
  \city{Stanford}
  \country{USA}}
\email{mlam4@stanford.edu}

\author{Joon Sung Park}
\affiliation{%
  \institution{Stanford University}
  \city{Stanford}
  \country{USA}}
\email{joonspk@stanford.edu}

\author{Kayur Patel}
\affiliation{%
  \institution{Apple Inc.}
  \city{Seattle}
  \country{USA}}
\email{kayur@apple.com}

\author{Jeffrey T. Hancock}
\affiliation{%
  \institution{Stanford University}
  \city{Stanford}
  \country{USA}}
\email{hancockj@stanford.edu}

\author{Tatsunori Hashimoto}
\affiliation{%
  \institution{Stanford University}
  \city{Stanford}
  \country{USA}}
\email{tatsu@cs.stanford.edu}

\author{Michael S. Bernstein}
\affiliation{%
  \institution{Stanford University}
  \city{Stanford}
  \country{USA}}
\email{msb@cs.stanford.edu}




\newcommand{\mlg}[1]{}
\newcommand{\msb}[1]{}
\newcommand{\msl}[1]{}
\newcommand{\thc}[1]{}

\renewcommand{\shortauthors}{M.L. Gordon, M.S. Lam, J.S. Park, K. Patel, J.T. Hancock, T. Hashimoto, M.S. Bernstein}

\newcommand{\na}{\textcolor{lightgray}{\textsc{n/a}}}

\begin{abstract}
Whose labels should a machine learning (ML) algorithm learn to emulate? 
For ML tasks ranging from online comment toxicity to misinformation detection to medical diagnosis, different groups in society may have irreconcilable disagreements about ground truth labels.
Supervised ML today resolves these label disagreements \textit{implicitly} using majority vote, which overrides minority groups' labels.
We introduce \textit{jury learning}, a supervised ML approach that resolves these disagreements \textit{explicitly} through the metaphor of a jury: defining which people or groups, in what proportion, determine the classifier's prediction.
For example, a jury learning model for online toxicity might centrally feature women and Black jurors, who are commonly targets of online harassment.
To enable jury learning, we contribute a deep learning architecture that models every annotator in a dataset, samples from annotators' models to populate the jury, then runs inference to classify.
Our architecture enables juries that dynamically adapt their composition, explore counterfactuals, and visualize dissent. 
A field evaluation finds that practitioners construct diverse juries that alter 14\% of classification outcomes.

\end{abstract}




\begin{teaserfigure}
  \includegraphics[width=\textwidth]{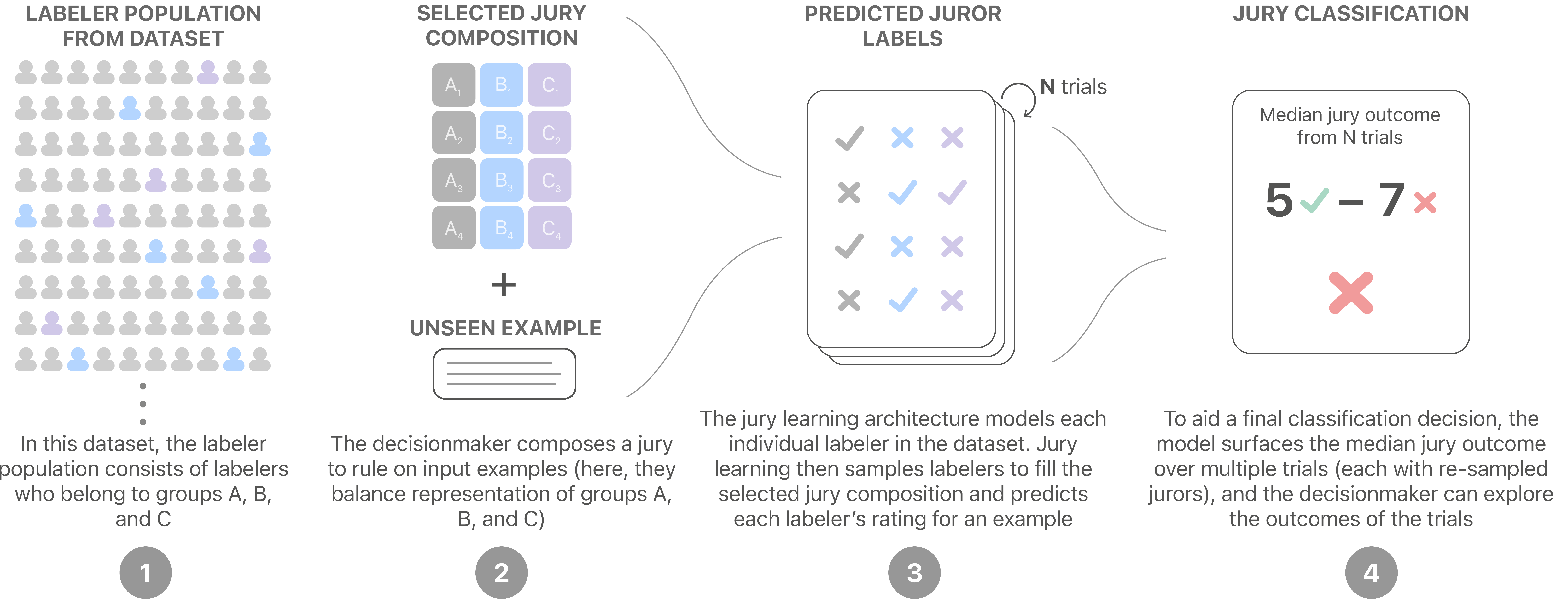}
  \caption{An overview of jury learning. (1) Given a dataset annotated by labelers from different groups, (2) the machine learning practitioner can compose a jury to rule on an unseen input example by allocating seats to labelers from the dataset with specified characteristics. (3) Then, the jury learning architecture models each individual labeler in the dataset, and performs \textit{N} trials in which it samples labelers as jurors to populate the specified jury composition and predicts each juror's decision for the example. (4) The system then outputs a median-of-means jury outcome alongside  jury outcome exploration visualizations that the decisionmaker can use to reach a classification decision.}
  \Description{This figure provides an overview of the Jury Learning approach. In the first section labeled “1,” we see a pool of many icons representing people (labelers), most of whom are colored grey, and a few are colored purple and blue; these represent three group identities. The subtitle of section 1 is “The labeler population consists of labelers who belong to groups A, B, and C”. In the second section labeled “2,” we see an icon representing a text blurb that is labeled “Unseen Example”, a plus sign, and a grid of 12 boxes labeled “Selected Jury Composition”. These 12 boxes representing juror slots are individually labeled A1, …, A4, B1, …, B4, C1, …, C4, where all A-boxes are grey, all B-boxes are purple, and all C-boxes are blue to represent that these juror slots map to each of the three group identities in equal proportions. The subtitle of section 2 is “The decisionmaker composes a jury to rule on the provided input example (here, they balance representation of groups A, B, and C). In the third section labeled “3,” there is an grid of check- and x-marks laid out in the same grid and colored to match the juror slots; there is a circular arrow labeled “N trials” and rectangles stacked behind the top rectangle to represent multiple trials. This graphic is labeled “Predicted Jury Member Labels” and the subtitle of section 3 is “For each trial, the jury learning model samples labelers to fill the selected jury composition and predicts each labeler’s rating for this input example.” In the fourth section labeled “4,” there is a box that contains text saying “Median jury outcome from N trials” that displays “5 check-marks vs. 7 x-marks” to indicate the jury outcome and an x-mark below to indicate the final verdict. This graphic is labeled “Jury Classification” and the subtitle of section 4 is “To aid a final classification decision, the model surfaces the median jury outcome from the trials, and the decisionmaker can explore the outcomes of the trials.”}
  \label{fig:teaser}
\end{teaserfigure}

\maketitle

\section{Introduction}

Whose voices---whose labels---should a machine learning algorithm learn to emulate?
In supervised machine learning today, the answers to these questions are often left \textit{implicit} in the data collection and training procedure. In a typical procedure, the practitioner pays multiple annotators to label each example~\cite{gray2019ghost}, then aggregates those labels via majority vote~\cite{lease2011quality,sheng2008get} into a single ground truth label~\cite{10.1145/3411764.3445402,suresh2019framework}.
The algorithm then trains on this aggregated ground truth, learning to predict ground truth labels that represent the largest group's point of view.

While this majoritarian~\cite{lijphart1999patterns} procedure succeeded for many early machine learning tasks, it now runs aground on tasks where there is substantial disagreement on what the correct label ought to be~\cite{gordon2021disagreement}.
Tasks with substantial disagreement are common in user-facing contexts, including classification of online comment toxicity~\cite{van2018challenges,georgakopoulos2018convolutional}, news misinformation~\cite{zhou2018fake, allen_arechar_pennycook_rand_2020}, and medical diagnosis~\cite{schaekermann2019understanding}. 
In these tasks, up to \textit{one third} of expert annotators disagree with each other when labeling an average example.
Properly accounting for labels from non-majority groups in a comment toxicity task, for example, reduces classifier performance from 0.95 ROC AUC---nearly solved---to a much less persuasive 0.73 ROC AUC~\cite{gordon2021disagreement}. This less persuasive number is indicative of the fact that it is impossible to create a classifier that makes every user happy---we have to make a choice.


Today's supervised learning approach, however, does not afford the technical or interactive tools necessary to resolve annotator disagreements through an \textit{explicit}, carefully considered choice. One response is to train the model to output a distribution across annotators rather than across classes~\cite{pavlick2019inherent,peterson2019human,liu2019learning,wang2019classification,zhang2017predicting}---e.g., ``40\% of annotators will say this comment is toxic, and 60\% will not.'' However, for an HCI researcher or practitioner who is designing a classifier that must make decisions in the face of disagreement, the quantity of interest is rarely just a question of how many people disagree, but one of \textit{who} disagrees and why~\cite{zhu2018value}. Reflective practices around dataset generation~\cite{gebru2018datasheets} can help specify whose voices a classifier should be designed to emulate during the dataset collection stage.
However, once a dataset has been collected and the resulting model trained, today's supervised learning pipeline does not afford the ability to reason over disagreement and then change a classifier's voices as tasks change or culture shifts. In most cases we lack even awareness of the need to do so: practitioners are typically unaware of whether stakeholders for a particular deployment or inference will disagree with a classifier's decisions, because they haven't modeled every annotator's or group's opinions. There remains a gap in providing algorithmic and interactive mechanisms that resolve the who, why, and decision rules of machine learning under societal disagreement.

In this paper we introduce \textit{jury learning}, a supervised learning architecture that closes this gap through the metaphor of a jury. Jury learning models every individual annotator in the dataset, enabling the practitioner to declaratively define which people or groups from the training dataset, in what proportion, should determine the classifier's prediction. The jury learning model architecture then predicts each juror's label and outputs the joint jury prediction to classify unseen examples. Rather than a typical machine learning classifier outputting a label of, for example, toxic or not toxic, a jury learning classifier might output a prediction such as, ``For this jury of six men and six women, which is split evenly between White, Hispanic, AAPI, and Black jurors, 58\% of the jury are predicted to agree that comment is toxic.'' 
Through jury learning, practitioners can define jury compositions that reflect stakeholders for the task, for example that the toxicity classifier should centrally feature women and Black jurors because they are commonly targets of online harassment~\cite{pewharassment2021,marshall2021algorithmic}.
The jury can articulate specific individuals, or any group-based annotation in the dataset (e.g., gender identity, political affiliation, racial identity).

To make a prediction on a new input, jury learning samples jurors from the practitioner's articulated jury composition, predicts each juror's response to the new input, then aggregates those responses into a final prediction.
Our jury learning exploratory interface (Figure~\ref{fig:system}) visualizes how each juror voted, enabling sensemaking about the nature of disagreement on an input or set of inputs.
This approach reconsiders the annotators who label training datasets not as inputs to an aggregation function but as a population of potential jurors.
To ensure that no groups are represented as singular and monolithic in their opinions, jury learning does not model groups but instead individual jurors. 
This model architecture enables visualizations that highlight where each sampled juror falls relative to the distribution of all annotators in that group. 

We contribute algorithms and visualizations that enable jury learning, then demonstrate them on a popular user-facing task of toxicity detection.
The core technical challenge: how do we achieve jury-based prediction from a dataset of similar size and scope as those already in use today, and without abandoning the architectures that make modern machine learning models highly performant?
We introduce a model architecture that combines state of the art natural language processing pipelines with techniques drawn from recent advances in deep learning based recommender systems~\cite{nguyen2020bertweet,wang2021dcn}.
This joint model architecture trains the algorithm to predict how every individual person in the training data would label previously unseen examples, much like a movie recommender system might model how a user would react to a movie---but with the added challenge that every inference is on an example previously unseen by anyone in the training data.
Our architecture enables this prediction task, and in addition enables visualization of the uncertainty underlying each decision---how many juries with the same composition would have ruled differently?---while highlighting differences between groups' predicted labels.
It also facilitates highly expressive jury-based algorithms, for example those that conditionally adapt the jury composition based on the relevant stakeholders for the input (e.g., populating with religious groups when the questionable comment is about religion, and political groups when the questionable comment concerns politics).
In addition, by adapting techniques from quadratic programming, we demonstrate that developers can understand how jury composition impacts classifier behavior through counterfactual juries: automatically identifying the smallest change to the jury composition that would reverse a decision.

In an evaluation, we test whether jury learning changes which groups influence classifications of a machine learning algorithm for toxicity.
Moderators of online communities (N=18) were asked to author juries for a comment toxicity classification task. We find that the resulting juries contain 2.9 times the representation of non-White jurors and 31.5 times the representation of non-binary jurors compared to those created implicitly by a large toxicity dataset~\cite{kumar2021designing}. This increased diversity in the jury composition changed the algorithm’s classifications on 14\% of items, reflecting the fact that jury learning captured those individual jurors’ views far better than a baseline, state of the art aggregated model (with an MAE of 0.62 versus 1.05).
We further find that our model architecture is \textit{more} accurate at predicting aggregate test set labels (MAE=0.27) than today's state of the art classifiers (MAE=0.41). This finding, which highlights the inherent instability of ground truth in the standard aggregate labeling approach, means that our model architecture both enables highly performant jury learning verdicts and also offers performance gains in the traditional aggregated task. Both of these are achieved by modeling each individual annotator whose opinions make up an aggregate label or jury verdict.

Taken together, this work contributes algorithms and interfaces for a machine learning architecture that makes explicit the selection of whose voice, with what weight, determines each prediction. We argue for this approach normatively, demonstrate its predictive accuracy, and produce evidence that practitioners' jury learning classifiers result in material changes in classifier behavior.

\section{Related Work}
In this section, we motivate jury learning through an integration of research in human-computer interaction---especially social computing---along with work in machine learning and AI fairness.


\subsection{Engaging stakeholders in algorithm design}
Our work draws on a critique of strict and unexamined \textit{majoritarianism} in governance~\cite{lijphart1999patterns}, which tends to exclude the viewpoints of minority groups~\cite{de1835democracy}. To protect minority rights, governance structures in practice typically include mechanisms that help avoid a tyranny of the majority (e.g., a bicameral structure~\cite{riker1992justification}). How should machine learning practitioners respond to this challenge? Jury learning presents one possible response, in which we introduce new levers that enable explicit control of how majorities are formed. In doing so, jury learning raises awareness of each potential majorities' consequences and encourages intentionality in their selection. We argue that in the hands of a well-intentioned actor, jury learning represents meaningful progress towards the problems that strict, unexamined majoritarianism can bring in machine learning. Doing so also opens opportunities for participatory and democractic approaches to jury selection.

Researchers in human-computer interaction and artificial intelligence have long articulated the need for algorithms to balance multiple stakeholders' needs, motivations, and interests, and to help achieve important collective goals~\cite{zhu2018value, moreau2019paradigm, costanza2020design, dobreski2018toward, smith2020keeping, yu2020keeping, abebe2020roles, bernstein2021esr}. One such thread, stemming from ethics in AI, focuses on ensuring fairness of outcomes. It demonstrates how machine learning training algorithms can enforce mathematical notions of individual~\cite{dwork2012fairness} and group~\cite{agarwal2018reductions,barocas2016big,hardt2016equality} fairness in classification tasks such as recidivism prediction. We build on advances in algorithmic fairness that help manage disparate \textit{outcomes}~\cite{mehrabi2021survey}, by contributing a technique that instead helps manage disparate \textit{beliefs}: whose labels we should be learning when there are irreconcilable disagreements among groups in society. For instance: in today's fairness approaches, the developer may normatively decide what a fair outcome looks like: e.g., comments submitted by Black users should be removed just as frequently as comments submitted by White users. Our work focuses on an orthogonal aspect of fairness: while disparate outcomes might focus on how many of these comments should be removed from different groups of users, we ask \textit{whose voices} should be involved in the decision of whether a comment should be removed.



Closer to our aims, a second thread of work proposes design guidelines and frameworks to help system designers ensure they are creating algorithms that reflect their stakeholders' values~\cite{zhu2018value, rahwan2018society, yu2020keeping}. These design processes argue for explicit inclusion of appropriate stakeholders in the design and evaluation of the algorithm. For instance, in WeBuildAI~\cite{lee2019webuildai}, stakeholders design their own models representing their beliefs, and then a larger algorithm uses each of these models as a single vote when making a decision for the group.
We agree that stakeholders' voices should be directly modeled in algorithmic systems.
We contribute a jury-based metaphor, along with a model architecture and algorithms designed to empower practitioners to explicitly resolve disagreements between stakeholders while retaining the performance of today's machine learning pipeline. 



In creating our approach, we draw on recent work that adopts a civics and governance metaphor for socio-technical design. Contested platform decisions can be made by juries of platform members, which can increase the perceived legitimacy of the decisions~\cite{fan2020digital, 10.1145/3134697}.
Platforms such as Facebook have recently engaged such models for setting decision-making precedent, as in their Oversight Board~\cite{klonick2019facebook}.
The PolicyKit toolkit demonstrates how such participatory processes can be encoded directly into the software that powers these platforms~\cite{zhang2020policykit}. 
Our work extends these metaphors to demonstrate their power in fully algorithmic environments as well, where they offer legitimacy and interpretability benefits.

\subsection{Disagreement, datasets, and machine learning}
Across tasks such as identifying toxic comments~\cite{van2018challenges,georgakopoulos2018convolutional}, bot accounts~\cite{wang2010detecting}, and misinformation~\cite{zhou2018fake}, researchers and platforms~\cite{kahn_2020} increasingly turn to machine learning to aid their efforts~\cite{gillespie2018custodians}. Specifically, these models are often trained using a supervised learning pipeline where we:
\begin{enumerate}
\item Collect a large dataset of individual beliefs, either generated through crowdsourcing services that ask several labelers to annotate each item according to policy and then aggregate the result into a single ground truth label (e.g.,~\cite{imagenet_cvpr09}), or similarly by asking and then aggregating experts (e.g.,~\cite{tucker2019crowdsourcing}).
\item Use those ground truth labels to train a model that produces either a discrete binary prediction or a continuous probabilistic prediction for any given example.
\end{enumerate}
For instance, in a Kaggle competition that received over 3,000 submissions, researchers were challenged to discover the best-performing architecture in a toxicity detection task~\cite{kaggle}. Facebook makes the vast majority of moderation decisions through classifiers~\cite{barrett2020moderates}, and YouTube does similarly~\cite{caplan2020tiered}. 

Classifiers typically speak with one voice, an aggregated pseudo-human that reflects the majority voice in the dataset they have been trained on~\cite{gebru2018datasheets, rogers2021changing, schaekermann2020ambiguity}. 
This majority-voice outcome can arise for two reasons: (1) majority vote aggregation of the raw crowdsourced annotations overrides minority viewpoints in generating ground truth, or (2) even if training data points are disaggregated, the training algorithm minimizes its loss function by predicting accurately for the opinions held by the largest group of people in the dataset.
Unfortunately, while this majority-voice approach to classification has been highly successful in many artificial intelligence (AI) tasks such as image classification~\cite{imagenet_cvpr09}, the results for many tasks in social computing and HCI remain problematic. 


One potential explanation for these problems may be that the voice a model has learned is not the right voice for every deployment, or even every inference within a deployment. To see how this might be true, we can examine annotator disagreement rates in today's datasets: for instance, in a toxicity task, over one third of annotators on average disagree with any toxic classification, even after accounting for label noise~\cite{gordon2021disagreement}; in a misinformation classification task, three professional fact checkers were unanimous on only half of URLs~\cite{allen_arechar_pennycook_rand_2020}. Across countries, content that was perceived as more or less harmful varies significantly~\cite{jiang2021understanding}.
Such disagreement indicates that there may be multiple competing voices, potentially representing different groups of people or sets of values. Indeed, a toxicity model tuned with a simple positive or negative offset (i.e. baseline) for each annotator achieves far more accurate per-annotator results than a standard classifier~\cite{kumar2021designing}. 

We build on research that aims to accurately capture the distribution of annotator opinions~\cite{chung2019efficient,dumitrache2017crowdsourcing,dumitrache2018capturing,dumitrache2015crowdsourcing,kairam2016parting,davani2021dealing}.
Given a dataset with individual annotator labels, machine learning researchers have begun training models to output a distribution of labels rather than a single class label, using loss functions such as cross-entropy compared to the distribution of annotators' labels~\cite{pavlick2019inherent,peterson2019human,liu2019learning,wang2019classification,zhang2017predicting}. 
While training models with cross-entropy loss acknowledges the existence of disagreement, it does not tell us \textit{who} disagrees or why, so we cannot readily act on it. An alternative approach, annotator-level modeling, has been shown to yield benefits to uncertainty estimation and majority vote prediction~\cite{davani2021dealing}. In this work, we introduce an annotator-level modeling architecture in the service of the decision rules underlying jury learning. As support for our approach, a Wizard of Oz study found that moving beyond raw distributions and towards AI-provided arguments for competing options resulted in users reviewing more contentious cases themselves~\cite{schaekermann2020ambiguity}.

Dataset documentation~\cite{gebru2018datasheets} and value-sensitive data collection practices~\cite{zhu2018value} can help specify whose voices a classifier should be designed to emulate. We build on these approaches in two ways. First, we provide an algorithm that makes clear when these voices disagree and provides tools to reflect on and re-weight whose voices are embedded in the model. From this perspective, our work innovates on this literature by directly modeling this information, allowing the machine learning practitioner to understand the nature of the disagreement and make explicit the representation that should resolve it. Second, our work addresses a practical reality of machine learning: while we cannot possibly have a universal set of voices that are appropriate for all models in a given task such as toxicity detection, existing approaches assume practitioners have the resources and motivation to collect new large-scale datasets every time the relevant stakeholders change. In reality, even in the rare cases in which practitioners have the required resources to collect new datasets, they are often unaware of the need to do so: we cannot know whether stakeholders will disagree with a classifier's decisions unless we've modeled every annotator's or group's opinions, leaving many practitioners unaware of the extent to which they are ignoring the opinions of certain annotators or groups. It is therefore not sufficient to have a procedure that requires that requires prior knowledge of the optimal annotator population at the outset. We contribute an approach that can model each relevant individual or group from a dataset similar in size and scope as those already collected today, so that practitioners can reason over and specify which of these individuals or groups their model should and should not reflect, iteratively and reflexively.

A large body of work in both HCI and machine learning discusses how improved dataset collection practices may result in more performant and ethical classifiers. Often, dataset authors instead strive for a goal of impartiality, so that data is supposedly ``unbiased''~\cite{scheuerman2021datasets}. To achieve such a goal, crowdsourcing researchers have proposed a number of methods that aim to resolve annotator disagreement either by making task designs clearer or relying on annotators to resolve disagreement among themselves~\cite{10.1145/3242587.3242598,manam2018wingit,schaekermann2018resolvable,chang2017revolt,dumitrache2015crowdsourcing}. 
However, for tasks such as those common in social computing contexts, much of the disagreement is likely irreducible~\cite{kairam2016parting,gordon2021disagreement,prabhakaran-etal-2021-releasing}, stemming from the socially contested nature of questions such as ``What does, and doesn't, cross the line into harassment?''. The above methods may help resolve some disagreement in these datasets, but until such an unlikely time as there is ever to be a global consensus on questions such as what constitutes harassment, classifiers must make decisions that represent some users' voices more than others'. Jury learning offers one approach to this decisionmaking.

An alternative approach is to retrain a model's single voice to represent a desired group~\cite{barbosa2019rehumanized, 10.1145/3394486.3409559, akhtar2021opinions}. If this decision can be made effectively up front, and the practitioner has the substantial budget and resources required to collect their own large-scale dataset, then a single data collection and training pipeline can succeed. Jury learning contributes an approach that allows real-time exploration and tuning of this population without requiring practitioners to collect new and far larger datasets, and makes stronger guarantees about whose voice is being represented in each specific inference.

\subsection{Interactive Machine Learning}
Our work draws on a recent thread of research integrating human-centered methods into machine learning systems. Interactive machine learning seeks methods to integrate human insight into the creation of ML models (e.g.,~\cite{amershi2014power,fails2003interactive}). One general thrust of such research is to aid the user in providing accurate and useful labels, so that the resulting model is performant~\cite{chang2017revolt}. Another line of work has sought to characterize best practices for designers and organizations developing such classifiers~\cite{amershi2019software,amershi2019guidelines}. Our work extends this literature, focusing on ameliorating issues that developers and product teams face in reasoning about their models and performance~\cite{patel2008examining}. 


A third line of works demonstrates that end users struggle to understand and reason about the resulting classifiers. Many are unaware of their existence~\cite{eslami2015always}, and many others hold informal folk theories as to how they operate~\cite{devito2018people}. In response, HCI researchers have engaged in algorithmic audits to help hold algorithmic designers accountable and make their decisions legible to end users~\cite{sandvig2014auditing}. Our work extends this literature, positioning classifiers as a reflection of many different voices, enabling control over that composition of voices, and enabling both practitioners and users to easily understand which voices are contributing to their models. 

\section{Jury Learning}
Machine learning often aims to emulate people's labels. Faced with annotator disagreement representing multiple competing voices, which people should we be emulating---whose training labels should a classifier use to make its decisions? We take the position that it is the machine learning practitioner's responsibility to make explicit normative decisions about whose voices their classifiers are reflecting in any given inference. In this section, we describe how we designed jury learning and our motivation in making each of these design decisions.

\subsection{Design goals}
We begin by considering today's approach. Many models return the class label or labels with the highest likelihood (e.g., label = `toxic', confidence .9). Some models instead predict the distribution of opinions over all annotators: for instance, that 60\% of annotators will label a comment as toxic, 30\% will label a comment as non-toxic, and 10\% will label as unsure.
How is a practitioner to act on this information?
If their goal is always to satisfy the largest number of annotators, the answer is easy.
However, there are many scenarios in which that is not---or should not---be the goal.
The practitioner may want to consider different voices (representing different values, experiences, or expertise) depending upon the situation.
Consider that a member of the LGBTQ+ community may be more informed about transphobic comments than the population at large. When a comment targets LGBTQ+ issues, or if a community is centered on supporting LGBTQ+ members, a practitioner may wish to more heavily weigh the opinion of these annotators.
Or consider that when doctors labeling MRI data disagree about a patient's diagnosis, the practitioner may wish to more heavily weigh opinions from doctors with a particular background or training. Or, it may be the case that the practitioner isn't initially sure who to side with, and so would like to reason over the different decisions that different annotators or groups of annotators would make.


It is, in theory, possible to achieve some of these goals using today's standard supervised learning pipeline. For instance, a practitioner deploying a classifier to the LGBTQ+ community could collect their own dataset, ensuring that a sufficient portion of annotators identify as LGBTQ+ so that disagreements are resolved by more heavily weighing opinions from LGBTQ+ annotators. In practice, however, such an approach fails to meet our goals. Datasets are expensive and difficult to collect, so practitioners often rely on existing datasets they did not collect, meaning they do not control how disagreements are resolved, and worse: \textit{do not even know that voices they care about are dissenting}. Without such knowledge, practitioners cannot reason over the different decisions that different annotators or groups of annotators would make. We require a different approach from today's standard supervised learning pipeline.


\subsection{Approach and interaction}
\textit{Jury learning} is a supervised learning approach that asks practitioners to specify whose voices their classifiers reflect, and in what proportion. To achieve this, jury learning models every individual annotator in a dataset, so that their model may serve as a potential juror. Practitioners then articulate a set of jurors that should be sampled from the groups or individuals in the annotators. That jury's labels determine the classifier's behavior.

For our purposes, we refer to a \textit{jury} as a bounded set of individuals whose opinions aggregate into a decision. These individuals are randomly sampled from the population of labelers based on the jury composition that the machine learning practitioner has articulated (e.g., six conservative jurors and six liberal jurors). Jury learning then algorithmically predicts how each of these twelve selected jurors would label the input, and then aggregates those responses into a decision. For many of our examples, we refer to a twelve person jury, which is the default jury size in the American legal system. However, the jury can be any size, if there are enough annotators in each group in the dataset to populate it.
If the task is regression rather than classification (e.g., a toxicity score rather than a binary toxic-or-not decision), the outcome is an average of jurors' predicted scores.

Jury learning enables the creation of many possible classifiers from a single dataset of labels, with the added requirement that the dataset contain information about each annotator for any group or voice that the practitioner wishes to include. For instance, the toxicity dataset we use as our example application domain~\cite{kumar2021designing} includes education, past work experience or qualifications, racial identity, gender identity, political affiliation, age, disability status. Practitioners specify jurors either individually, or using any group membership criteria available in the dataset. If a practitioner selects a juror using group-based data, we demonstrate where that juror fits within the full distribution of all annotators within that group, ensuring that no group is represented as monolithic. Practitioners can interactively explore different jury compositions, gaining an understanding of the consequences of each composition that they try: how are specific annotators or groups of annotators differing in their labels?

%

\begin{figure*}
  \includegraphics[width=\textwidth]{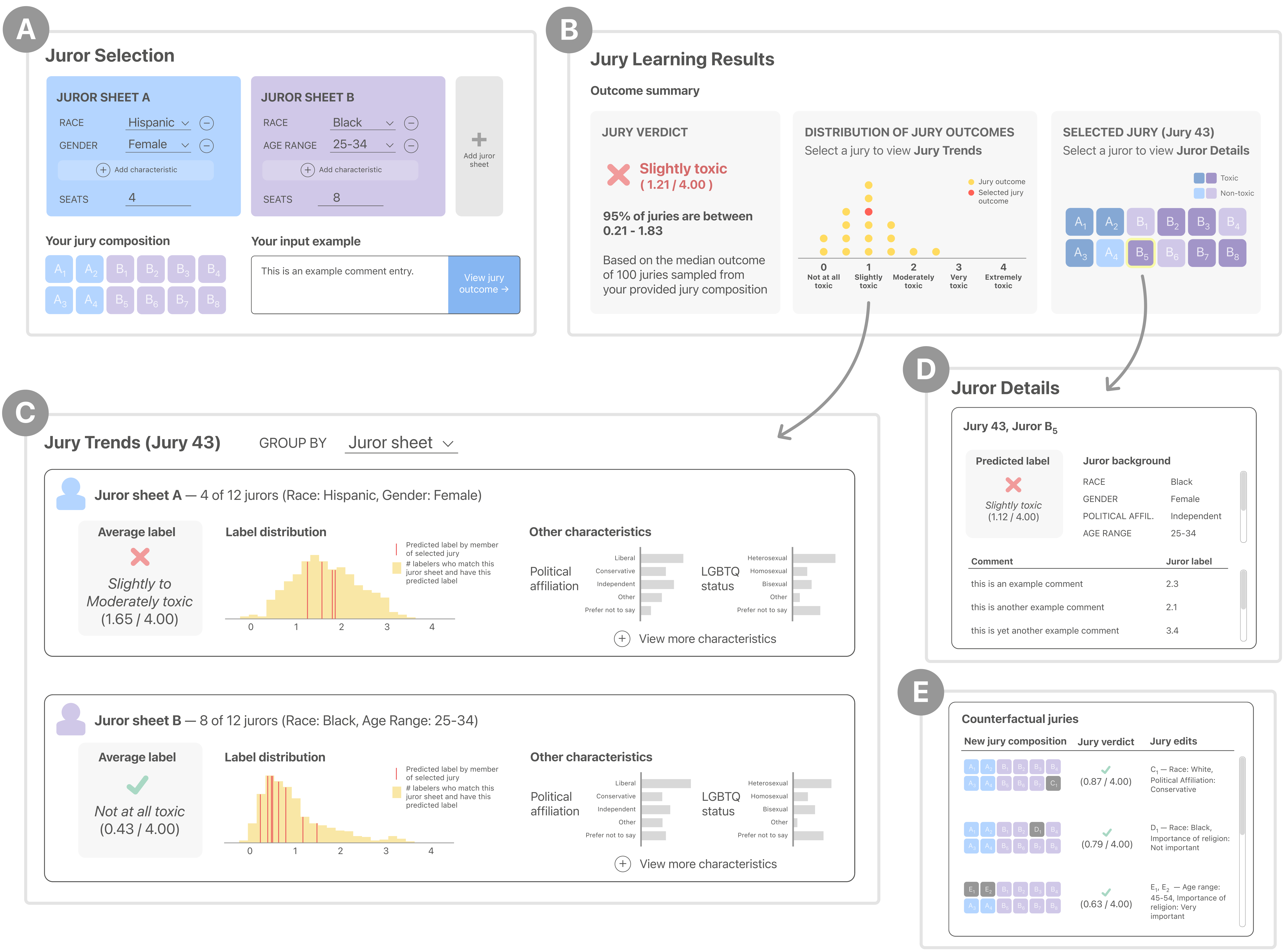}
  \caption{System Overview. (A) In the Jury Selection portion of the system, the user can create juror sheets to populate their jury composition and can provide one or more input examples to evaluate. (B) Then, the system outputs the Jury Learning Results section where they can view a summary of the jury verdict based on a median-of-means estimator of jury outcomes. Here, they can view the full distribution of jury outcomes, select individual juries to view trends, and inspect individual jurors on a jury. (C) When a user selects a jury, the Jury Trends section is updated. There, they can group by different fields like the juror sheet, decision label, or other demographic attributes to understand patterns in the labels from this jury and contextualize them with respect to the larger population. (D) When a user selects a particular juror, the Juror Details view opens, and they can inspect the predicted label for the juror, the background of this juror, and the juror's annotations. (E) Users can also inspect counterfactual juries that would result in the opposite verdict.}
  \Description{This figure shows the key elements of the Jury Learning system. In the first portion labeled “A,” there is an excerpt of the UI entitled “Juror Selection.” There are two boxes labeled “Juror Sheet A” and “Juror Sheet B”; Juror Sheet A specifies Race to be Hispanic and Gender to be Female and indicates 4 juror slots; Juror Sheet B specifies Race to be Black and Age range to be 25-34 and indicates 8 juror slots; there are buttons below the specified attributes to add additional characteristics to each juror sheet, and there’s a button on the right to add more juror sheets. Below the juror sheets is a grid of boxes labeled “Your jury composition that is colored and labeled to match the two displayed juror sheets. There is an text box to the right labeled “Your input example” where a user could enter a comment, and there’s a submission button to the right of the box labeled “View jury outcome.” In the section labeled “B,” there’s an excerpt of the UI entitled “Jury Learning Results” and subtitled “Outcome Summary.” There are three interface elements: (1) a Jury Verdict box that displays the text “Slightly Toxic (1.21/4.00). 95\% of juries are between 0.21-1.83. Based on the median outcome of 100 juries sampled from your provided jury composition.”, (2) a Distribution of Jury Outcomes box that displays a dot plot visualizing the scores of all of the juries, and (3) a Selected Jury box that shows the grid of jurors that is colored to indicate which jurors voted “toxic” versus “non-toxic.” In the section labeled “C,” We see a UI portion labeled “Jury Trends (Jury 43).” Here, there are two boxes; one for Juror Sheet A and one for Juror Sheet B. Each box has an Average Label section that shows the average label of jurors sampled from that juror sheet, a Label Distribution section that shows a histogram of all labels from jurors who match the juror sheet (with the sampled jurors’ labels marked by red lines), and an Other Characteristics section where there are bar charts indicating the distribution of labelers’ characteristics such as Political affiliation and LGBTQ status. In the section labeled “D,” we see a UI element labeled “Juror Details” with a box that displays the predicted label, juror background, and juror annotations for a selected juror from the jury. In the section labeled “E,” there is a UI element labeled “Counterfactual juries” where there is a table showing the new jury composition in one column (represented by colored juror slots), the jury verdict in the next column, and jury edits in the last column (a text description of the different attributes that the new jurors have).}
  \label{fig:system}
\end{figure*}

Figure~\ref{fig:system} displays the jury learning interface for our example application domain of toxicity detection.
In Figure~\ref{fig:system}(A), practitioners specify a jury composition by assigning a \textit{juror sheet} to each of twelve empty \textit{juror slots}. A juror sheet defines the characteristics of the annotator who will fill the juror slot. A simple juror sheet may specify only one characteristic, such as a juror identifying as Black, while a more complex juror sheet may articulate an intersectional identity, such as a juror identifying as a Black LGBTQ+ woman. The possible characteristics are dictated by the provided dataset: the set of identities must be broad enough to reflect the relevant stakeholders~\cite{scheuerman2021revisiting}.
If a characteristic is better captured through open ended text boxes than categories~\cite{scheuerman2021revisiting}, the practitioner could explore individual people in the dataset and select a subset for inclusion. The jury composition can be defined interactively via a web interface. The interface also allows the machine learning practitioner to explore different jury compositions and how each might react to different inputs.

The machine learning practitioner can then apply their jury to an input or set of inputs. Given an input to predict, jury learning makes a prediction for that input for every annotator. It then takes a step not possible when convening real-world juries, but possible with jury learning: it convenes many parallel iterations of the jury, by repeatedly resampling a large number of juries that match the jury specification. Each jury may contain different jurors (annotators from the dataset), and the model will predict different responses to the input for each juror based on that juror's training data in the dataset. The interface disallows selecting any groups with an insufficient number of jurors in the dataset to complete the resampling procedure without replacement, directing practitioners to collect more data for the particular group. 

The system then responds with a jury verdict for the input: the single, final decision of the median jury on that input, shown in Figure~\ref{fig:system}(B). To identify a verdict, the system samples a set of individual jurors filling the jury specification, predicts each juror's decision, and then determines the aggregate verdict taken as a majority vote (for classification) or mean (for regression). The default decision is calculated as the median jury decision from the set of sampled juries matching the jury specification. This median-of-means estimator~\cite{lecue2020robust}---the median of the mean juror responses across juries---produces an estimate that is robust both to variance within groups and to potential juror-level modeling errors by the AI. In particular, this approach is resistant to the model being wrong about any small number of jurors, though less effective for systematic errors that may impact most or all jurors.

\begin{figure}
  \centering
\begin{minted}[fontsize=\scriptsize,samepage]{python}
jury = [
    {
        'jurors': 4,
        'gender_identity': 'female',
    },
    {
        'jurors': 4,
        'gender_identity': 'nonbinary',
    },
    {
        'jurors': 4,
        'gender_identity': 'male',
    }
]
\end{minted}
  \captionof{figure}{ The jury definition can be passed as a Python dictionary object.}
  \label{fig:input}
  \Description{A Python dictionary object consists of a list of dictionaries; each dictionary has the keys “jurors” (mapped to an integer value) and “gender_identity” (mapped to a string like “female” or “nonbinary”).}
\end{figure}


\begin{figure}
  \centering
\begin{minted}[fontsize=\scriptsize,samepage]{python}
result = {
    'verdict': 'toxic',
    'votes': {
        'toxic': .67, # 8 of 12 jurors voted 'toxic'
        'nontoxic' .33, # 4 of 12 jurors voted 'nontoxic'
    },
    'jurors': [
        {
            'juror_id': 1023,
            'gender_identity': 'female',
            'racial_identity': 'White',
            'political_affiliation': 'liberal',
            'vote': 'toxic',
        },
        {
            'juror_id': 2342,
            'gender_identity': 'female',
            'racial_identity': 'South Asian',
            'political_affiliation': 'conservative',
            'vote': 'nontoxic',
        },
        ...
    ],
    'population': {
        'toxic': .85, # 85% of sampled juries voted 'toxic'
        'nontoxic': .15, # 15% of sampled juries voted 'nontoxic'
    }
}
\end{minted}
  \captionof{figure}{The response likewise is returned as a Python dictionary object.}
  \label{fig:ouput}
  \label{lst:representation_examples}
  \Description{A Python dictionary object consists of the keys “verdict” (mapped to a string “toxic”), “votes” (mapped to a dictionary with keys “toxic” and “non-toxic” that in turn map to proportions of jurors in the median jury who voted each way), “jurors” (mapped to a list of dictionaries that each list juror information like juror ID, gender identity, racial identity, political affiliation, and vote), and “population” (mapped to a dictionary with keys “toxic” and “non-toxic” that in turn map to the proportion of sampled juries that voted each way)}
\end{figure}

The approach also results in a direct measure of uncertainty: how often the outcome changed across the jury samples. For example, the system might communicate that 85\% of juries matching the specification resulted in a ``toxic'' label, and 15\% of juries resulted in a ``not toxic'' label. Or, for a regression task, it might communicate a histogram distribution of jury decisions, as shown via the histogram in Figure~\ref{fig:system}(B).



Because the system returns a specific jury, the system can visualize each juror in context of the group from which they were sampled (Figure~\ref{fig:system}(C)). This contextual information helps the machine learning practitioner better understand the behavior of the jurors chosen for their jury. Specifically, for each juror, we make available all of their annotations and all associated background information that is present in the dataset. This visualization also helps make clear that different members of a group may vote differently, and that despite this individual variation, the overall jury outcome may be stable. In addition, our approach grounds the jurors as individual people with specific characteristics and enables other explainability-related information, such as highlighting how the juror labeled similar inputs in the training data or providing their specific modeling error rate over all of their test examples \ref{fig:system}(D)).

The system is interactive to encourage better sensemaking, but it also provides a code layer for automated systems. The jury definition can be passed as a Python dictionary object, as in Figure~\ref{fig:input}. The response likewise is returned as a Python dictionary object, as in Figure~\ref{fig:ouput}.

\subsection{Example scenario}
Saanvi has created an online news-sharing social network, and wants to create a classifier to detect any instances of personal attacks on the platform. She finds a popular, publicly available large-scale dataset, trains a model using the traditional supervised learning pipeline, and deploys it to her community. The classifier takes as input the text of a comment, and returns a ``toxic'' or ``not toxic'' label. Unfortunately, Saanvi soon begins to notice that both she and many members of her community often disagree with the decisions this classifier makes. 
Saanvi suspects that perhaps her classifier isn't making decisions in ways that reflect the voices in her community.

Saanvi's dataset contains characteristics about each annotator, so she switches from the traditional classifier to a classifier created through jury learning. First, Saanvi explores different jury configurations to confirm any group-based differences that she expects to see, inputting comments and exploring how the jurors in each group respond. She confirms that men are more likely to rate borderline comments as not toxic, but notes that there are many women on her platform. By exploring, she also observes that seniors find more comments to be toxic, and 18--35 year olds find fewer comments to be toxic. Saanvi begins by constructing a jury that she believes better represents the members and values of her community. She deploys a private test of it, and notices a significant improvement: the classifier's decisions start making a lot more sense to her and her community members. Saanvi then begins a participatory process to bring in stakeholders from her community, allow them to test different jury configurations, and agree upon a jury to use on their platform.

Saanvi and the other stakeholders observe that their intuitions of the proper jury composition change based on which groups might be targeted in that post: that when a news article is about women's issues, they want more women on the jury; when a news article concerns LGBTQ+ rights, they want more jurors identifying as LGBTQ+; when an article is about a Black woman, they want more Black women on the jury. So, they agree to dynamically allocate four seats on the jury to the appropriate group based on the news category that the post is shared in (e.g., four women for news articles shared in the womens' rights category).

Saanvi exports the model and puts it into private testing on her server, where its predictions are not yet shown to users. She and the group of stakeholders continue to monitor its behavior.
Eventually, as they build trust in the algorithm, they begin to use it to prioritize comments for human moderators on the platform.



\section{Technical approach}



Jury learning requires that we predict how each individual annotator would label an unseen example. A jury outcome is then an aggregation of the jurors' (annotators') individual classifications.

Enabling the broadest set of applications also requires an approach that can make such predictions from a dataset of similar size and structure to those already in use when training supervised standard classifiers: a labeled dataset with a few annotators labeling each item and each annotator labeling a few items, such as those commonly acquired from crowdsourcing services. The only additional assumption we make is that any characteristic used to select jurors (e.g., gender identity) must exist for each juror. This is achievable by adding a small survey when an annotator begins labeling examples. In what follows, we describe our model architecture for jury learning. While we focus our description on natural language processing tasks (specifically, toxicity detection), the high-level architecture is general and can apply to any inputs that allow content embeddings (e.g., images via Resnet~\cite{szegedy2017inception}, screens via Screen2Vec~\cite{li2021screen2vec}, or text via BERT~\cite{devlin2018bert}).

We base our model architecture on the insight that, in trying to predict how each annotator would label an unseen example, we share part our goal with the aim of today's recommender systems. Like recommender systems, we must not only perform well over a range of inputs, but also over a range of individuals. Like recommender systems, we expect that different opinions between annotators can often be partly explained by explicit categorical information about the groups that each annotator belongs to or identify with, but are also partly unique to a particular annotator or explained by unobserved latent factors~\cite{app7121211}. In other words, much like how Netflix might develop a model to predict individual users' opinions on films, our jury-based model will predict individual labelers' perspectives on new inputs.

Unlike Netflix, however, all of the inputs to our model will be unseen examples (or, in recommender systems language, all examples suffer from the \textit{cold start problem}), meaning that they have never been seen by any annotators in our training set. This is a standard assumption in any classification task, but not a typical assumption of most recommender systems, which often rely heavily on an item's existing annotations to inform what other users will think of it. We require an approach that relies entirely on an input's featurization: by taking an input and embedding it, we can predict an annotator's label by comparing this input to similar examples they have already annotated. This means that today's hybrid deep learning recommender systems for natural language input, which typically train their own item embeddings~\cite{wang2021personalized}, are insufficient. We propose a model architecture that jointly trains a content model for classification tasks (such as from BERT) alongside a deep recommender system. By combining deep recommender systems' ability to model individuals' opinions with modern pre-trained deep learning models' classification task performance, our architecture takes full advantage of the strengths of each.

\begin{figure*}
  \includegraphics[width=.8\textwidth]{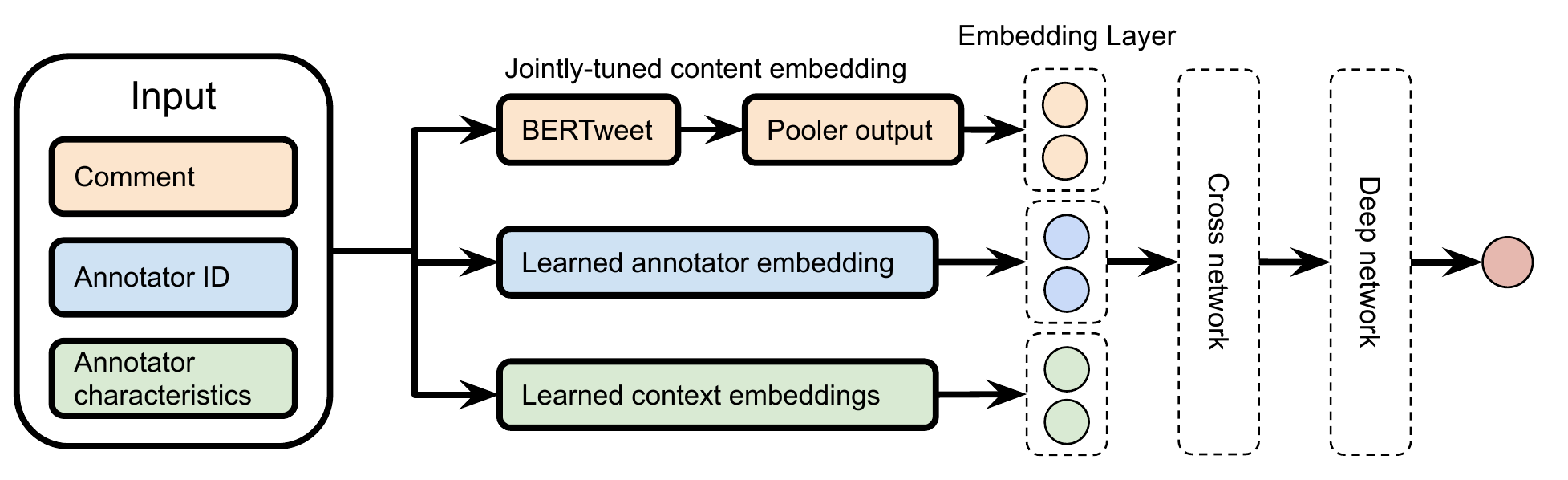}
  \caption{We introduce a model architecture that jointly fine tunes the practitioner's existing content-based classifier alongside a Deep \& Cross Network recommender system.}
  \label{fig:model_architecture}
  \Description{This model architecture diagram displays a box labeled “Input” that contains boxes labeled “Comment,” “Annotator ID,” and “Annotator characteristics”. This box has arrow that branch out into three rows of boxes. The first row has a BERTweet box with an arrow to a Pooler output box, and both of those boxes are labeled “Jointly-tuned content embedding”. The second row has a Learned annotator embedding box. The third row has a Learned context embeddings box. Each of these rows feeds into a section labeled “Embedding Layer” that is represented by rectangles containing circles within them. This layer has an arrow feeding into a Cross network box that then feeds into a Deep network box which feeds into a single circle that represents the output classification.}
\end{figure*}

For our recommender system architecture, we select a Deep \& Cross Network (DCN)~\cite{wang2021dcn}. DCNs were designed for web-scale collaborative filtering applications in which data are mostly categorical, leading to a large and sparse feature space. While DCNs were created for classic recommender system tasks, our insight is that a modified DCN architecture is strong fit for jury learning. A typical DCN involves three sets of embeddings: content, annotator, and group. The content embedding enables prediction on previously unseen items by mapping those items into a shared space. The group embeddings make use of the data from all annotators who belong to each group, helping overcome sparsity in the dataset. The annotator embedding ensures that the model learns when each annotator differs from the groups they belong to. The DCN learns to combine these embeddings to predict each individual annotator's reaction to an example: the embeddings are concatenated into an input layer, then fed into a cross network containing multiple cross layers that model explicit feature interactions, and then combined with a deep network that models implicit feature interactions~\cite{wang2021dcn}. We modify the DCN architecture to jointly train (or more precisely in the case of a pre-trained models, jointly fine tune) a pre-trained BERT-based model, using its pooler output as the content embeddings. Figure~\ref{fig:model_architecture} displays a high-level view of our end to end model architecture.

\subsection{Implementation for toxicity detection}
Having described our high-level approach and architecture, which can be applied to a wide range of tasks, we now turn to the specific task we use to demonstrate jury learning in this paper: toxicity detection.

\subsubsection{Dataset description}
We train our model using a publicly available balanced dataset~\cite{kumar2021designing} in which 107,620 social media comments were labeled by five annotators each, from a pool of 17,280 unique annotators. This dataset was collected to understand how user expectations for what constitutes toxic content differ across demographics, beliefs, and personal experiences. Each annotator labeled a minimum of 20 comments, with a small fraction labeling more than 20. Each annotator contained categorical information noting their self-identified gender, race, education, political affiliation, age, whether they're a parent, and whether they consider religion an important part of their lives. Annotators were asked to rate each social media comment's toxicity on a scale from 0 to 4, with 0 being non-toxic, 1 being slightly toxic, and 4 extremely toxic. If we binarize this task, with a rating of $< 1$ indicating non-toxic and $>= 1$ indicating toxic, we find that annotators in this dataset disagree with each other 35.9\% of the time.

\subsubsection{Training}
We use TensorFlow Recommenders (TFRS) as the basis of our implementation. TFRS natively supports DCNs. We use Huggingface's Tensorflow API to instantiate BERTweet ( a large-scale language model pre-trained on English Tweets, released by NVIDIA~\cite{nguyen2020bertweet}) as the pre-trained content embeddings within our recommender system. We adapt the model to the task by performing an initial fine-tuning step on a large-scale toxicity dataset released by Jigsaw~\cite{kaggle}.

Initially, we co-train all the model's components together: we fine tune the pre-trained large language model, and we train from randomly initialized values for the annotator embedding, group embeddings, and the DCN. However, while BERT-based models have been shown to quickly overfit after fine tuning for a few epochs, our newly initialized components can benefit from a longer training procedure. We therefore co-train the entire model for two epochs, freeze the large language model, and continue training the remainder of the model for 8 epochs. Further epochs did not noticeably improve the model's performance. We used the Adam optimizer and Mean Squared Error as our loss function.

We trained our model on one machine with one NVIDIA Titan XP GPU. The majority of the Titan XP's memory is taken up by BERTweet, so most of the DCN itself is stored in the machine's memory during training. We chose standard hyperparameters used when fine tuning BERT-based models: we used learning rate of $2e-5$, a batch size of 16, and a maximum length of 128 tokens. We set our DCN-specific hyperparameters as follows: we set a constant embedding dimension of 32, a three-layer cross network of size 768, three dense layers of size 768, and a output dense layer of size 1. We selected these sizes and the number of training epochs after performing a small grid search.


\section{Extensions}
The architecture of jury learning directly affords new decision-making and interpretability techniques that are not available with traditional algorithms. Here we overview two such techniques that we have implemented.

\subsection{Conditional juries}
We might desire different forms of expertise depending on the decision at hand. For example, CHI's peer review process identifies jurors (reviewers) who differ for each paper under review, based on the content of the paper. Likewise, civil society organizations convene different groups of stakeholders depending on who their decisions might impact. In the context of AIs, for example, classifying misogynistic comments may call for a jury with a larger representation of women, whereas classifying racist comments may call for a jury with a larger representation of minoritized racial groups.

While the default jury learning algorithm focuses on a simple metaphor of a stable jury composition that is used for all decisions, jury composition can be \textit{conditional} on the item being classified. A simple code conditional might adapt the jury composition:

\begin{minted}[fontsize=\scriptsize,breaklines]{python}

# Define a default six of the twelve jury members, allowing the other six to vary based on the context
default_jurors = [ ... ]

# select the other six jurors based on context
conditional_jurors = []
if '#metoo' in tweet:
    conditional_jurors = [
        {
            'jurors': 6,
            'gender_identity': 'female',
        }
    ]
elif '#blm' in tweet:
    conditional_jurors = [
        {
            'jurors': 6,
            'racial_identity': 'Black',
        }
    ]
elif ... # additional conditions and conditional jurors

# combine the default six jurors with the six jurors who have been selected for this context
jury = default_jurors + conditional_jurors
\end{minted}

Alternatively, approaches such as clustering or topic modeling might be appropriate:

\begin{minted}[fontsize=\scriptsize,breaklines]{python}

jury = [ ... ] # default jury

# embedding_distance calculates the comment's cosine similarity to comments that contain a given term
if embedding_distance('#blm', tweet) < .05:
    jury = ... # jury composition for Black Lives Matter topics
elif embedding_distance('#metoo', tweet) < .05:
    jury = ... # jury composition for MeToo topics
elif embedding_distance('vaccination', tweet) < .05:
    jury = ... # jury composition for vaccination topics
elif ... # additional conditional juries for specific topics that the community cares about
\end{minted}



\subsection{Counterfactual juries}
When a jury decides a given comment to be non-toxic, it naturally gives rise to the question: what jury composition, if any, \textit{would} find the comment to be toxic? How different would the jury need to have been to flip the outcome? Jury learning enables this interaction to search for a \textit{counterfactual jury}, by automatically identifying the minimal change to the jury composition that would result in a different outcome than the current jury (Figure~\ref{fig:system}(E)).

Within the jury learning framework, we frame the search for the counterfactual jury as an optimization problem: flip the classification by making the smallest edit possible to the current jury composition.
Formally, we can define this as a quadratic program, solvable via convex optimization. Consider that we have $K$ different annotators or groups of annotators, and we have a prediction $s_k$ associated with each. We set the size of our jury, $n_{jurors}$, to 12, meaning that we must assign a value in $\{0 \hdots K\}$ to each of the 12 juror slots.
To represent a jury composition, we define a jury allocation vector $p$ of length $K$. Each index of $p$ refers to an annotator or group in $K$, and the value at each index refers to the number of jurors from the corresponding annotator or group. The jury allocation vector should therefore should sum to $n_{jurors}$.
The classification decision we consider is a threshold on a jury's average prediction, which we define as $v_p = \frac{\sum_k {p_k} {s_k}}{n_{jurors}}$. The final classification is based on whether $v_p > 1$.
The problem of identifying a counterfactual jury is now equivalent to a quadratic program. If the current decision is in the negative class $v_p \leq 1$, then the counterfactual jury that flips this decision is defined as the solution to the following optimization problem
\begin{align*}
  \min_{p^* \in \mathbb{Z}^+} \sum_k &(p_k-p^*_k)^2 \\
  &s.t. \\
  \sum_k p^*_k = n_{jurors} \quad\text{ and }\quad   &v_{p^*} > 1  \quad\text{ and }\quad p^*_k \geq 0. \\
\end{align*}
This optimization problem can then be solved by off-the-shelf optimization solvers.


Counterfactual juries can serve as a useful interpretability lens, aiding the community in understanding how dependent the classification outcome was on the jury composition.

\section{Model evaluation}



Taking our example application of toxicity detection, we evaluate the performance of our proposed model architecture at two levels:
\begin{enumerate}
    \item How accurate are individual juror predictions?
    \item How accurate are the final predictions produced by a jury?
\end{enumerate}

The most important question to test with jury learning is whether the learning algorithm correctly estimates what jurors' opinions are on previously unseen data.
Recommender systems make predictions across different individuals by identifying commonalities among annotators and borrowing information.
Without an approach that sometimes borrows information, building a jury learning system would require acquiring a large dataset from each group, including each intersectional identity group, which is often infeasible. However, any machine learning approach that borrows information also brings a risk: it is possible to borrow \textit{too much} information, particularly when we have less data from a specific group or annotator.
So, our evaluation seeks to test whether the approach is correctly estimate each juror's labels.

\subsection{Individual juror performance}

\begin{table*}[hbt!]
\centering
\small
\resizebox{\textwidth}{!}{\begin{tabular}[t]{l|lllllll}
\toprule
& Full test set & Asian & Black & Hispanic & White & Male & Female \\
\midrule
Number unique annotators & 11262 & 817 & 1774 & 424 & 9087 & 6077 & 6985 \\
\midrule
MAE: Baseline aggregated model & 0.90 & 0.83 & 1.12 & 0.87 & 0.87 & 0.94 & 0.86 \\
\midrule
MAE: Jury learning model & 0.61 & 0.62 & 0.65 & 0.57 & 0.60 & 0.61 & 0.60 \\
\bottomrule
\toprule
& Liberal & Independent & Conservative & Asian+Female+Liberal & Hispanic+Male+Conservative \\
\midrule
Number unique annotators & 5388 & 3764 & 3687 & 206 & 54 \\
\midrule
MAE: Baseline aggregated model & 0.86 & 0.86 & 1.01 & 0.84 & 0.96 \\
\midrule
MAE: Jury learning model & 0.60 & 0.58 & 0.65 & 0.62 & 0.64 \\
\bottomrule
\end{tabular}}
\caption{Performance against individual annotator's test labels for three models: today's standard state-of-the-art aggregate approach (which is annotator-agnostic, and makes one prediction per example), a group-specific version of our proposed architecture, and the full version of our proposed architecture. The standard aggregated model's performance varies substantially between groups. For instance, it achieves an MAE of 0.83 for Asian annotators and 1.12 for Black annotators, a performance decrease of 35.0\%. By comparison, we find that our model does show differences between groups, but with far smaller magnitudes. It achieves an MAE of 0.62 for Asian annotators and 0.65 for Black annotators, a performance decrease of 4.9\%.
}
\label{table:deepak}
\end{table*}

\subsubsection{Performance versus a standard classifier}
We first demonstrate that jury learning is substantially more accurate in predicting individual annotator responses when compared to a baseline state-of-the-art, annotator-agnostic classifier.


To create a state of the art baseline model, we fine tune BERTweet on the toxicity dataset using the same standard hyperparameters we used to fine tune BERTweet within the jury learning algorithm. As the toxicity dataset provides a regression task in (0, 4), we report the Mean Absolute Error (MAE), comparing each individual annotator's predicted response to their observed response. We design the test set for this evaluation so that all of the comments were never seen by the annotators in the training set.
This more challenging prediction task reflects the expected usage of our model, as discussed earlier. Our test set contains 5,000 comments and 24,545 annotations.

We find that our model achieves an MAE of 0.61, and the baseline model achieves an MAE of 0.90. This large improvement is not necessarily surprising: our architecture is the only one that makes use of information about individual annotators. This result demonstrates that our model was able to learn a substantial amount of useful information about each annotator or their groups; if jury learning had learned nothing about either an individual or groups, then its predictions would simply match those of a standard state of the art classifier trained on aggregated labels, which makes one prediction per example.

\subsubsection{Performance versus a group-based classifier}
The above performance gains could either have come from learning about individual annotators, the groups they belong to, or both. Our goal with jury learning is to ensure that models are not solely reliant on group membership; we would also like our model learn about how individual annotators diverge within their groups. We therefore now ask: how performant is our model at predicting individual annotators' responses to an example, compared to an ablation of our model that only knows about group membership? If our model performs better using both annotator and group information than solely group information, it has learned specific information about annotators.

To create a group-specific classifier, we train a model using our proposed architecture with one change: we remove annotator IDs as a feature, meaning that our model can only rely on group-based and content-based features. We find that this model achieves an MAE of 0.81. This score is an improvement over the baseline aggregated classifier's 0.90, indicating that our model learned useful information from group-based features. However, our full individual+group model's MAE of 0.61 is a substantial improvement over both, indicating that our full architecture is reliant on both group and individual annotator features.


\subsubsection{Is our model more performant for some groups than others?}
A recommender-like prediction system may implicitly group `similar' individuals together (due to its low-rank inductive bias), leading to some unique individual and intersectional perspectives being erased. Such issues could give practitioners false confidence that they are accounting for intersectional opinions, decrease public confidence (as individuals can verify predictions are incorrect for them), and lead to decision systems that are worse than the status quo. In particular, this issue could arise for smaller groups where our model may need to borrow more information. Addressing this issue requires first understanding its extent. We therefore now ask: is performance consistent across groups of varying sizes?

This section is not an exhaustive study of intersectional identities in our dataset, which would be infeasible to report. Rather, we focus on three of the most salient group-based categories in our dataset (race, gender, and political affiliation), shown in Table \ref{table:deepak}. As illustrative examples, we also report results for two intersectional identities. 

We first note that the baseline aggregated model's performance varies substantially between groups. For instance, it achieves an MAE of 0.83 for Asian annotators and a far worse 1.12 for Black annotators, a performance decrease of 35.0\%. By comparison, we find that while our model does show differences between groups, but it does so with far smaller magnitudes. It achieves an MAE of 0.62 for Asian annotators and 0.65 for Black annotators. 

\subsection{Jury-level performance}
Having shown that our architecture can model individual annotators, we now turn to jury level predictions. Ultimately, these are the most important predictions that our model makes. We ask: how performant is our model at predicting a jury's verdict?

To evaluate jury-level predictions, we'd like to compare the predicted final value produced by a jury against an observed final value produced by the same jury. Ideally, we would use comments in our test set that have been labeled by at least 12 annotators, and treat those 12 annotators as a de-facto jury.

While our dataset does not contain comments labeled by twelve annotators, it does contain a subset that were labeled by ten annotators. We rely on this small subset to get a close approximation (though likely a slightly pessimistic estimate) of the MAE of a 12-member jury. We define the \textit{observed verdict} as the mean observed annotation over all 10 annotators, who serve as the de-facto jury. We define the \textit{predicted verdict} as the mean of our model's individual predictions for those same ten annotators. Over 550 10-annotator juries, we find that our model produces a jury-level MAE of 0.27.

We have shown in the previous sections that jury learning is very effective when the annotators of interest are different from the distribution of annotators in the original dataset (e.g. intersectional identities). However, we show a surprising result: jury learning is more effective than the current aggregate prediction approach \textit{even when the annotator distribution is the same as that of the dataset}. We find that the above baseline model produces an MAE of 0.41 over aggregate test labels, notably worse than our model’s 0.27. These gains are due to the fact that these examples are annotated by a small group of 10 annotators where the identity of each annotator has a strong influence on the observed verdict, and jury learning can make predictions that account for the identity of these jurors.

\section{User Evaluation}
Having demonstrated the technical efficacy of our jury learning architecture in making annotator-level and jury-level inferences, we then sought to evaluate jury learning in the hands of real-world stakeholders in the content moderation setting. Our study aimed to answer the following questions:
\begin{itemize}
    \item Q1: What jury compositions do participants select? How diverse are the selected jury compositions with respect to the implicit jury compositions embedded in the original dataset?
    \item Q2: Do participant-specified juries result in different prediction outcomes than those produced by a standard classifier?
\end{itemize}

To answer these questions, we targeted our study towards two audiences in the context of our focal task of toxicity classification: content moderators and platform users. Given their expertise in making policy decisions that are tailored to the needs of particular online communities, content moderators are the population most likely to directly utilize our system.

In the supplementary materials, we replicate this study with everyday platform users who are not involved in content moderation and who might not currently feel that they have a voice in this decision-making, and we also report on survey instruments measuring the perceived legitimacy (willingness to grant deference and authority) of jury learning compared to traditional algorithms.


\subsection{Study design}
We conducted an online study that consisted of a Qualtrics survey with two main components: (1) a jury composition section where participants were asked to design a jury for an online community and answered several short-answer follow-up questions, and (2) a moderation algorithm legitimacy section where they answered questions to assess their perceptions of the legitimacy of a current moderation algorithm and the proposed jury algorithm. To ground the survey in a concrete scenario, we framed all of the questions in terms of a hypothetical online social media platform called YourPlatform that is planning to use algorithmic approaches as a major component of its content moderation strategy. At the start of the survey, we provided a detailed explanation of a \textit{current algorithm} (a standard machine learning classifier trained on human labels using majority vote label aggregation) and a \textit{jury algorithm} (an instantiation of our jury learning approach) and explained that YourPlatform was considering using one of these methods.

For the jury composition task, we displayed one of 5 possible comment sets (generated by random samples from our comment toxicity dataset~\cite{kumar2021designing} stratified by toxicity severity and labeler disagreement) to exemplify the type of content they would need to moderate on YourPlatform. Participants were then shown a simplified jury composition input form that allowed them to allocate 12-person jury slots using three demographic attributes: (1) \textit{gender} (Female, Male, Non-binary, Other), (2) \textit{race} (Black or African American, White, Asian, Hispanic, American Indian or Alaska Native, Native Hawaiian or Pacific Islander, Other) and (3) \textit{political affiliation} (Conservative, Liberal, Independent, Other). While our approach can accommodate as many categorical values as are associated with labelers, we selected this limited set of axes because they are common demographic attributes that capture a fair amount of variation among users and that were relevant to the topics of the comment sets.
Further details on our study procedure and the full survey contents are found in our supplementary materials.

\subsection{Participant recruitment}
For our content moderator study, we recruited participants who serve as moderators for Discord or Reddit. A member of our research team recruited Discord moderators via a server where many Discord moderators gather to discuss issues around moderation and recruited Reddit moderators of major subreddits via individual solicitation. Due to their domain expertise and relative scarcity, we offered content moderators \$40.00 to complete our 30 to 45-minute survey. In total, 18 content moderators participated in our study. These participants moderate on a variety of platforms (17 on Discord, 5 on Reddit, and 2 on Twitch; some participants moderate across multiple platforms and communities). Based on self-reported demographics, we had 9 participants of age 18-24 and 9 participants of age 25-34; we had 4 women, 9 men, 4 non-binary individuals, and 1 participant who did not disclose their gender; we had 12 White, 2 Asian, and 3 multi-racial participants (1 participant did not disclose their racial identity).

\subsection{Analysis approach}
To analyze our results, for each available demographic attribute value, we compared its representation in participant juries against its corresponding \textit{current algorithm implicit jury} representation. The \textit{current algorithm implicit jury} represents the proportion of each demographic group in the original dataset. For each demographic attribute, we calculated the proportion of labelers for each comment who possessed that attribute and computed the average of these per-item proportions across the dataset. These proportions were normalized among the subset of demographic attributes that we selected for this study. The current algorithm implicit jury determined through this process---the annotators in the training data for the current algorithm---is 74\% White (see red lines on Figure~\ref{fig:mod_jury_diversity}).

In addition, both survey sections had open-response questions. The goal of our analysis here was to summarize high-level themes that emerged from these responses, so a member of the research team read through all responses multiple times to generate a set of themes using qualitative open coding~\cite{32_Charmaz}, then coded comments according to these themes.

As a post-study analysis step, we took participants' jury compositions and performed inference with our jury learning algorithm to compare the jury-based outcome with that of a standard ML algorithm.

\subsection{Results: Jury composition diversity (Q1)}
First, we examined the jury compositions designed by our participants. We had a total of eighteen moderators who completed our survey, of which we were able to analyze sixteen.\footnote{We exclude two of the moderators' jury composition results: while these participants demonstrated an accurate understanding of the current algorithm and jury algorithm (and thus have valid moderation legitimacy responses), they utilized the ``Other'' fields to mean ``any'' or ``null,'' but this field was defined to map to jurors who explicitly self-identified with ``Other'' for these attributes. Since these responses are not directly comparable, they have been excluded from the quantitative jury composition analysis.} All possible values for all three attributes were utilized in the study, and participants constructed diverse juries with a mean of 5.7 unique race values (SD=0.85), 3.1 unique gender values (SD=0.56), and 3.4 unique political affiliation values (SD=0.61). This diversity involved the explicit inclusion of non-majority identities (here, defined as values other than White for race, Male or Female for gender, and Liberal or Conservative for political affiliation): on average, participants created juries with 10.31 individuals (SD=1.26) who had one or more non-majority attributes; participants created juries with on average 3.88 individuals (SD=1.76) who had two or more non-majority attributes (e.g., Black and Non-binary).

\begin{figure*}
  \includegraphics[width=\linewidth]{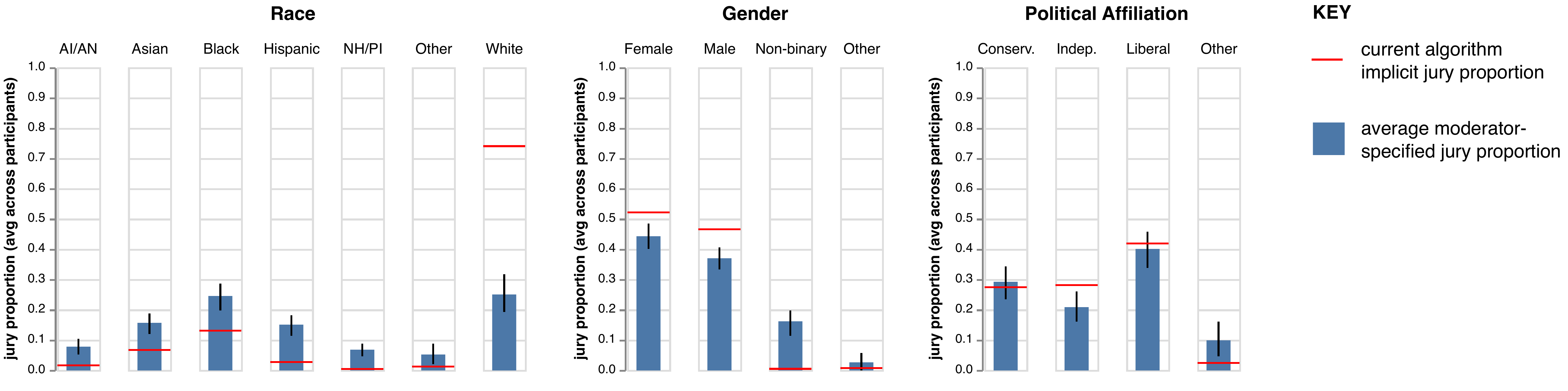}
  \caption{Jury composition results ($N=16$). While there are sizeable disparities in group representation in the current algorithm implicit jury (denoted with red lines), the moderator-specified juries generally achieve greater diversity (raising representation for groups with the lowest red lines and lessening the gap in representation among groups).}
  \label{fig:mod_jury_diversity}
  \Description{The jury composition results are summarized here with three bar charts. In the chart entitled “Race,” there are bars for all seven racial identities used in this study. The red line indicating the current algorithm implicit jury proportion for White labelers is much higher (at 0.74) than all the others, which are at most 0.13. The blue bars indicating the average moderator-specified jury proportion are higher than the red line for all racial identities other than White, which is substantially lower at 0.25. In the chart entitled “Gender,” there are again bars for all four gender identities used in this study. The red implicit jury line is around or below 0.5 for Female and Male jurors and very close to 0 for Non-binary and Other jurors. The blue bar representing the moderator-specified juries are lower than the red line for Female and Male, much higher than the red line for Non-binary, and a bit above the red line for Other. In the chart entitled “Political Affiliation,” there are again bars for all four political affiliations used in the study. The red lines for Conservative, Independent, and Liberal labelers are all moderately high at around 0.3-0.4 and the Other line is near 0. The blue bars representing moderator-specified jurors are lower for Independent, similar height for Conservative and Liberal, and higher for Other political affiliations.}
\end{figure*}

\begin{table*}[t]
\centering
\footnotesize
\begin{tabular}{llrl}
\toprule
            Attribute &                                 Value &  t-statistic &   p-value \\
\midrule
                 Race &                               White** &       -14.79 & < 0.001 \\
                 Race &                               Asian** &         4.82 & < 0.001 \\
                 Race &           Black or African American** &         4.84 & < 0.001 \\
                 Race &                            Hispanic** &         6.50 & < 0.001 \\
                 Race &    American Indian or Alaska Native** &         4.35 &  < 0.001 \\
                 Race & Native Hawaiian or Pacific Islander** &         5.60 & < 0.001 \\
                 Race &                                Other* &         2.16 &     < 0.05 \\
\midrule
               Gender &                                Male** &        -5.18 & < 0.001 \\
               Gender &                              Female** &        -3.51 & < 0.01 \\
               Gender &                                 Other &         1.13 &     \textit{n.s.} \\
               Gender &                          Non-binary** &         7.06 & < 0.001 \\
\midrule
Political affiliation &                               Liberal &        -0.53 &     \textit{n.s.} \\
Political affiliation &                          Independent* &        -2.69 &     < 0.05 \\
Political affiliation &                                Other* &         2.43 &     < 0.05 \\
Political affiliation &                          Conservative &         0.59 &     \textit{n.s.} \\
\bottomrule
\end{tabular}
\caption{Jury composition one-sample t-test results ($N=16$). Values denoted with double-asterisks (**) are significant with $p < 0.01$; values denoted with a single asterisk (*) are significant with $p < 0.05$. Most notably, we observe strongly significant increases in the representation of non-White jurors, strongly significant increases in the representation of Non-binary jurors and corresponding strongly significant decreases in the representation of Male and Female jurors.}
\label{table:mod_jury_stats}
\end{table*}

We then compared the diversity of the moderator-designed juries relative to the diversity of the current algorithm implicit jury we defined earlier.
As summarized in Figure \ref{fig:mod_jury_diversity}, we observed that for all three demographic attributes, participants juries achieved greater diversity than the current algorithm implicit jury. We performed one-sample t-tests comparing the representation of demographic attribute values between the current algorithm implicit jury and the participant jury and report the results in Table \ref{table:mod_jury_stats}. For racial identity, we observed strongly significantly decreases in the representation of White jurors ($p < 0.001$) and strongly significant increases ($p < 0.001$) in representation for all non-White race attribute values except for the ``Other'' category, where we still saw a significant increase in representation ($p < 0.05$); participants' juries contained 2.9 times the representation of non-White jurors than the current algorithm implicit jury. For gender identity, we observed a strongly significant reduction in both male and female jurors and a strongly significant increase in the representation of non-binary jurors ($p < 0.001$) with 31.5 times the representation of non-binary jurors compared to the current algorithm implicit jury. Finally, for political affiliation, we observed a significant decrease ($p < 0.05$) in the representation of Independents (who were oversampled in the original dataset) and a significant increase in the representation of other political affiliations.

\begin{table*}[!htb]
      \centering
      \footnotesize
        \begin{tabular}{p{5cm}c|p{3cm}c|p{5.5cm}c}
        \toprule
        Jury composition approach &  Count & Anticipated outcomes &  Count & Justifications for increasing/decreasing voice&  Count\\
        \midrule
         {Diversity, equal representation, fairness} & 13
         & Better capturing views of minority groups & 8
         & Extent to which <group> has relevant experience or knowledge about the issues at hand & 7 \\
         {Prioritizing groups targeted in sample comments} & 10
         & Increase in number of comments rated as toxic  & 6
         & Extent to which <group> is marginalized or has experienced historical harms & 6 \\
         {Increasing representation of minority groups} & 4
         &  &
         & Extent to which <group> has been targeted by the sample comments & 4 \\
         {Decreasing representation of groups that may cause harm to minority groups} & 1
         &  &
         & Extent to which <group> is expected to view as toxic the content that other groups would find toxic & 3 \\
        \bottomrule
        \end{tabular}
        \caption{In a field study, we asked participants ($N=16$) open response questions about their approach to composing juries, the outcomes they anticipated, and justifications for their jury composition decisions. We manually coded their responses to identify themes. The count column indicates the number of participants who mentioned each theme. A majority of participants aimed to prioritize diversity and equal representation of juror attributes, and the majority took special care to increase representation of groups who were targeted in the provided comment set.}
        \label{table:mod_jury_comp_qs}
\end{table*}

Our qualitative coding shed light on the reasons underlying participants' jury composition decisions. As summarized in Table \ref{table:mod_jury_comp_qs}, a vast majority of participants aimed to prioritize diversity and equal representation of juror attributes, and the majority took special care to increase representation of groups who were targeted in the provided comment set. When asked to envision how outcomes of the jury algorithm might differ from those of the current algorithm, many participants felt that it would better capture the views of minority groups and would increase the number of comments rated as toxic. Finally, when explaining which groups had more or less voice in their jury composition, many users stated that they based their decision on whether certain groups had relevant experience with the comment topic or whether certain groups had been historically marginalized or underrepresented.

\subsection{Results: Jury prediction outcomes (Q2)}

\subsubsection{How many classification outcomes flip between toxic and non-toxic?}
Having established that participants composed a diverse selection of juries, we now ask: do these participant-specified juries result in different prediction outcomes than those produced by a standard classifier? As our standard baseline classifier, we use the same state of the art BERTweet-based classifier defined earlier in the Model Evaluation section.

We first aim to establish that jury learning effectively models the individual jurors selected by participants. We therefore perform a disaggregated analysis in which we randomly sample jurors for each of the diverse, participant-provided jury composition 100 times. We then compute an MAE over all the comments labeled by all selected jurors. We find that jury learning decreases the average error of these diverse participant-provided juror’s opinions by 41\% when compared to the predictions from our baseline aggregated model, from an MAE of 1.05 to 0.62. 

We then focus on the final predictions produced by jury learning, computed through a median-of-means estimator over 100 resampled juries. We compare these predictions to the predictions from our baseline classifier. To determine whether a jury's prediction caused a toxicity decision to change, we binarize the final regression values found from our median-of-means estimator such that a value $< 1$, indicates non-toxic, and $\geq 1$, corresponding to a value ``slightly toxic'' or greater in the annotation scheme, indicates toxic. We remove a small number of juror sheets (mean: 4\%) because because the participant requested more jurors from an intersectional identity than available in the original dataset.

Over the 16 moderator-provided juries, we find that a mean of \textit{13.6\%} of decisions flip, with a standard deviation of 4.1\% across moderators. This result suggests that a meaningful number of classifications can change between an off-the-shelf classifier and a jury learning classifier customized for the community.

\subsubsection{Do diverse juries flip divisive comments?}
Having established that the diverse juries provided by participants cause toxicity predictions to flip, we now investigate \textit{which} comments are flipping. Specifically, we ask whether the comments that flip tend to be more divisive among annotators than the comments that do not flip. To make this determination, we compute an annotator disagreement rate for each comment in the test set. We find the annotator disagreement rate by randomly sampling pairs of annotations for the same comment, and computing the percentage of the time that these pairs disagree with each other. Across all comments that participants' proposed juries cause to flip, the annotator disagreement rate is 46.4\%. A two-proportion z-test shows this to be a significant increase over the 37.2\% disagreement rate for comments that these juries did not flip ($z = 2.89, p < .01$). This result indicates that jury learning has the biggest impact on comments that are the most divisive.

\section{Discussion}
In this section, we reflect on the contributions, limitations and future opportunities of our approach. We reflect on how designers and product teams might use it in practice. Finally, we reflect on the ethical considerations of our approach.
\subsection{Implications for design}
How do we build artificial intelligence systems that reflect our values? Values are often diverse and heterogeneous across individuals. While the raw datasets that most ML systems rely on are made up of individuals, today’s approaches to building machine learning classifiers typically abstract the individuals out of the pipeline. They view differences among annotators as label noise, rather than as genuine differences of opinion that practitioners need to understand and account for. Jury learning is an attempt to re-think the machine learning pipeline so that practitioners make explicit value judgements about the voices that their classifiers should reflect. We believe, and our evaluation suggests, that practitioners and researchers who make these decisions explicitly will include greater diversity than typical models today. Our approach centers individuals at each stage of the pipeline rather than abstracting or aggregating them as in today’s ML approaches.
\subsubsection{A new lens for ML interpretability}
Today, approaches for machine learning interpretability typically base their explanations around properties of the item in question, aiming to communicate how an item's features or content led the model to make its decision. Our approach affords a new, complementary lens to machine learning interpretability, in which we aim to explain a model's prediction as a function of the properties of its annotators.

Consider an activist whose social media posts are removed by an AI. They might rightfully wonder if their posts were moderated because the annotators that trained the moderation model had different political views. Such information is currently completely hidden, making it difficult for this activist to trust the outcomes of automated moderation systems. In contrast, jury learning enables new interpretable methods for users to interrogate which groups' opinions are being listened to, which groups' opinions are not, and for what kinds of inputs. Does it weigh men's voices more than women's in its training data? Does this amplify bias for some topics? Jury learning could empower end users to call for greater representation. More broadly, jury learning offers a new way for users and decision makers to communicate and debate normative decisions about whose perspectives should be included.




\subsection{Ethical considerations}
Compared to today's implicit procedure for selecting a classifier's voice, our explicit approach introduces its own ethical issues and trade-offs.




\subsubsection{Making fair and transparent decisions}
How do we eradicate harmful biases in machine learning? Existing approaches in the machine learning fairness literature largely take the training data as a given, and then enforce statistical constraints that can introduce notions of fairness on the resulting model's output (e.g., that a model's decisions must be equitable across genders). In other words, the existing fairness literature starts from the assumption that the underlying statistical correlations in the world are flawed, and that they must be corrected through post-hoc adjustments of decisions that were learned from a flawed world. However, these solutions are ultimately band-aids to a problematic input pipeline. A useful distinction is to consider different forms of justice. We can think of jury learning as a form of \textit{procedural justice}. We do not claim to guarantee the fairness of outcomes, but instead we make claims around the correctness of the process.
 \mlg{would love people's thoughts: should we expand on this? If so, how?}

Our work instead takes the position that it is sometimes more desirable or tractable to select specific people whose voices should be emulated. This position comes with its own set of challenges. While jury learning empowers and normatively encourages practitioners to think carefully about whose voices their models represent, it does not inherently enforce notions of fairness. Jury learning can be used to beneficially select the most important voices to a practitioner, or to equitably represent a diverse set of groups. Jury learning can also be used to unintentionally or deliberately make biased decisions that may cause harm. A practitioner could purposely exclude a relevant group's voice, or could unintentionally include a harmful voice. If, for instance, a practitioner unintentionally or intentionally selects racist jurors, then the resulting model will be racist.

However, unlike fairness approaches that focus on outcomes, the jury learning approach can make use of tools from the human-computer interaction and social sciences literature that provide established and effective levers to recruit, train, and socialize people such that a practitioner can overcome these challenges and achieve the jury composition that they want. We argue that, if the options are to make decisions by enforcing post-hoc constraints on the decisions learned from large and somewhat random datasets, or the jury learning approach of explicitly selecting people who make decisions, it is often better to go with the latter. In doing so, we can entrust decision-making to the most relevant, qualified people for any task or situation.

Beyond the juror selection considerations above, we also advocate for transparent juries. Even if jury learning leads to increases in diversity, jury learning is unlikely to dramatically re-order the existing power structures within sociotechnical systems. Rather, the aim of jury learning is to ensure that decision-making regarding issues of power, in particular whose voices are represented in classification tasks, is made explicit and transparent. We therefore propose that any organization deploying a jury-based classifier make their jury composition transparent to relevant stakeholders. In doing so, jury learning enables a new set of conversations between practitioners and stakeholders about precisely whose voices a classifier is emulating, the implications of emulating those voices, and the ability to explore and implement different sets of voices. Such conversations could be considered akin to a \textit{Batson challenge}, a process in the US legal system in which stakeholders to a case can argue against the removal of particular jurors on impermissible ground. To that end, we also suggest that practitioners make their instantiation of our jury learning interactive interface publicly available as a sandbox so that anyone can understand how different juries might make different decisions.

\subsubsection{Addressing the ecological fallacy}
Our aim with jury learning is to help practitioners recognize and integrate annotator disagreement in the classifier pipeline. To achieve this, we ask practitioners to create a jury that specifies the individuals or groups their classifiers should emulate. One approach to creating such a classifier might have been to simply model each group as a singular representative voice, akin to personas in traditional HCI methods. However, such an approach would promote an ecological fallacy because it does not demonstrate the extent to which annotators within a group disagree with each other. Our approach instead models individual annotators, enabling tools that inform practitioners about disagreement within groups. The amount of this disagreement depends upon the extent to which the group identities selected by the practitioner can explain disagreement between annotators.

Another risk arises from the requirement that many machine learning tasks produce a single decision. To make this decision, we must take a position that resolves any disagreement: specifically, we use a median-of-means approach that takes the median jury after randomly sampling 100 juries that match the practitioner's jury composition, ignoring ones that might be outliers. Thus, our system still presents an opportunity promote the ecological fallacy. To ensure that practitioners are aware of this risk, our interactive interface clearly communicates that each jury composition can have many different instantiations, and that a jury's verdict may change depending upon which jurors happened to be selected. Further, we promptly display visualizations that contextualize each individual juror within their larger group, demonstrating where they fall within the distribution of other annotators that may have been chosen in their stead. Finally, as mentioned in our system description, the interface disallows selecting any groups with an insufficient number of annotators in the dataset to complete the resampling procedure without replacement, directing practitioners to collect more data for the particular group.



\subsubsection{Accurate representation}
As with any machine learning system, our approach is only as good as the labels provided to it, and only as good as the model's ability to learn from these labels. If a dataset does not accurately represent the views of its annotators, or does not accurately convey each annotator's group memberships, then our model will emulate those inaccuracies.
Users of our system must therefore follow best practices when collecting their datasets. For instance, the dataset we used to demonstrate jury learning relies on self-identifications, which brings its own tradeoffs when compared to an approach that attributes identity characteristics to participants.

Further, no current model architecture can perfectly emulate the annotators it was trained on. The high stakes nature of social computing settings means that there can be substantial harm from misrepresenting minority perspectives. Good crowdsourcing practices should therefore be paired with participatory methods for auditing the models produced by jury learning, and any performance metrics should be split out by group to ensure that the model's performance is equitable across groups. Future work should also develop new techniques based on robust machine learning to ensure that models are trained to explicitly optimize for performance across all subpopulations rather than on average~\cite{hashimoto2018fairness}.


\subsubsection{Abdication of responsibility}
One risk of the jury learning approach is that it may provide a mechanism for platforms to both avoid taking broad policy stances and also evade blame for content moderation decisions. This stems from two aspects of its present design that remain open-ended: (1) the choice of the decisionmaker who wields the jury learning tool to make content moderation decisions, and (2) the meta-policy by which the jury learning outputs are incorporated into an end-to-end content moderation system (answering questions like: \textit{what circumstances do and don’t warrant the creation of a new jury? How do we weight the jury outcomes against other algorithmic tools' outcomes in a standard, principled way? How should we balance the jury outcome against the opinion of a content moderator? How do we select what comments should be sent to a jury?}). Ultimately, the organizations deploying classifiers are responsible for the decisions their classifier makes, and should still be held accountable for them.

\subsubsection{Annotator privacy}
Faithful and accurate representation of jurors potentially requires information collection about the private views and attributes of jurors. Factors such as sexual orientation are highly private, but can be a key part of creating a jury with diverse perspectives. Data recovery and record linkage attacks mean that such information could potentially be leaked to an adversary. Balancing the rights of jurors to privacy with the accountability and transparency benefits of leveraging juror demographics is a challenging open question. Future work in jury learning should investigate methods to disclose potential privacy harms to annotators. For instance, disclosure may require that, when collecting new datasets, we make clear to labelers the possibility that these attributes may be recoverable.
Future work should also draw on approaches for differential privacy in AI~\cite{Abadi_2016} to help ensure us that individuals or rare demographic attributes are not rediscoverable.

\subsection{Limitations and future work}
As with any machine learning approach, there are several limitations and future directions worth discussing:



\subsubsection{Domains}
In this paper, we demonstrated jury learning using a single application domain: toxicity detection. However, our approach is designed to work for any task in which there is annotator disagreement, a dataset denoting each annotator's relevant group memberships, and an existing classification model that produces high quality embeddings for each item. In particular, we hope future work will investigate using jury learning for medical decision making and design tasks, which may rely on different perspectives. For instance: a doctor making use of a model to help them decide between different treatment options might benefit if their model's decisions were based on a jury that reflects a particular patient's preferences in quality of life trade-offs. Or an amateur designer making use of an AI-based tool for poster design might benefit from the ability to create juries reflecting different design sensibilities or artistic schools of thought.

\subsubsection{Jury metaphor}
Jury learning loosely draws on a metaphor of juries in the US legal system, but we do not intend this rhetorical device to indicate a complete isomorphism. Rather, jury learning draws on two specific aspects of juries: the notion of moving from a single decision maker to a group of voting decision makers, and the idea of some sort of juror selection process.

Juries in the US legal system are the sites of complex social behaviors facilitated through an intricate legal apparatus~\cite{hastie2013inside}. These behaviors yield benefits and challenges to justice (for instance, group polarization~\cite{sunstein1999law}) and are not the focus of our system. For instance, jury learning does not draw on the deliberative nature of juries, which has been the subject of decades of study in legal literature~\cite{devine2001jury}. Jury learning's approach to juror selection also contrasts with the approach taken in the US legal system. Jury learning empowers practitioners and end-users to select their own jury composition. In the US legal system, jury selection is not in hands of single individual, but rather jurors are selected through a process in which stakeholders argue to determine its composition. As discussed above in our ethical considerations section, a stakeholder-centered selection process may sometimes be useful in jury learning, and existing work in the HCI literature~\cite{lee2014crowdsourcing,lee2019webuildai} demonstrates how such a process could be put into practice within our system.


\subsubsection{Group identifiers}
To demonstrate jury learning, we relied on an existing dataset that provided group membership information for each annotator. This dataset happened to focus on collecting this information for socio-demographic groups. One limitation to note is that the choice to use categories here has consequences. For instance, non-binary individuals find gender dropdown forms problematic unless they include appropriate nonbinary options and an open text box for description when appropriate~\cite{scheuerman2021revisiting}. One approach to creating inclusive interfaces in this respect is to ensure that all relevant options are represented in the jury interface. Another would be to allow the practitioner to explore the set of people who used the open-ended textbox and select a subset of them for inclusion as possible jurors.




Finally, our approach currently relies on datasets that include explicit information about the groups that each annotator belongs to. Future work should investigate unsupervised approaches to finding different voices within datasets~\cite{kairam2016parting}, potentially rendering the jury learning approach possible with any existing dataset.



\subsection{Positionality statement}
The authors represent backgrounds ranging from computer science (HCI, machine learning) to media psychology. We acknowledge critical arguments making thoughtful cases for removing AI from socio-technical systems, as well as arguments substantially increasing human control, oversight and audits of them. Our ideological commitment in this paper is to situations where improvement rather than outright removal of the AI is the appropriate mitigation strategy. We also acknowledge our shaping by the North American normative commitment to decisions being made by a jury of peers. Historically, juries have been sites of both progressive and regressive decision-making. Finally, we recognize that the term ``toxic'' is non-specific and often used as a catch-all term for a variety of forms of content that people do not wish to see online. In order to be consistent with the process used to collect this dataset, we draw upon this use of the term ``toxic.''

\section{Conclusion}
Machine learning often means learning to imitate people. So whose voices--whose labels--does a machine learning algorithm learn to imitate? Faced with endemic disagreement in user-facing tasks, we have to make a choice. But today's supervised learning pipelines typically abstract individual people out of the pipeline, treating people as abstractions or aggregated pseudo-humans. As a result, we lack the ability to reason over \textit{who} disagrees and why. Jury learning is an attempt to bridge the realities of machine learning with the realities of contested tasks. Our approach enables practitioners to make explicit value judgements that inform how models resolve disagreement. If successful, we hope that this approach will help developers make more informed and intentional decisions about creating and deploying classifiers in these contexts.

\begin{acks}
We thank Joseph Seering for his contributions to our user study. We thank Jane E, Harmanpreet Kaur, Danaë Metaxa, Ranjay Krishna, Matthew Joerke and James Landay for insightful discussions, feedback, and support. We thank Deepak Kumar for providing our toxic content dataset. We thank the reviewers for their helpful comments and suggestions. Mitchell L. Gordon was supported by the Apple Scholars in AI/ML PhD fellowship. This work was supported by the Computer History Museum, Patrick J. McGovern Foundation, and the Stanford Institute for Human-Centered Artificial Intelligence.
\end{acks}


\bibliographystyle{ACM-Reference-Format}
\bibliography{sample-base}


\end{document}